\documentclass[preprint2]{aastex}
\newcommand{\etal}{{\em et~al.}}
\newcommand{\ie}{{\em i.e.}}
\newcommand{\Rsun}{R_\odot}
\begin{document}
\title{An Upper Limit on the Temporal Variations of the Solar Interior
       Stratification} 
\author{A. Eff-Darwich\altaffilmark{1,2,5} and S.G. Korzennik\altaffilmark{1}}
\author{S.J. Jim\'enez-Reyes\altaffilmark{2,3}}
\author{F. P\'erez Hern\'andez\altaffilmark{2,4}}

\affil{$^1$Harvard-Smithsonian Center for Astrophysics, 60 Garden St., 
       Cambridge, MA, 02138 MS16}
\affil{$^2$Instituto de Astrof\'\i sica de Canarias,
       C/ V\'\i a L\'actea s/n, Tenerife, 38205, Spain}
\affil{$^3$High Altitude Observatory,  
       National Center for Atmospheric Research, 
       1850 Table Mesa Dr.
       Boulder, Colorado 80307}
\affil{$^4$Departamento de Astrof\'\i sica,
       Universidad de La Laguna, Tenerife, Spain}
\altaffiltext{5}{present address: Department of Soil Sciences and Geology, 
                 University of La Laguna, Tenerife, 38205, Spain}
\begin{abstract}
 We have analyzed changes in the acoustic oscillation eigenfrequencies
measured over the past 7 years by the GONG, MDI and LOWL instruments. The
observations span the period from 1994 to 2001 that corresponds to half a
solar cycle, from minimum to maximum solar activity.

  These data were inverted to look for a signature of the activity cycle on the
solar stratification. A one-dimensional structure inversion was carried out to
map the temporal variation of the radial distribution of the sound speed at
the boundary between the radiative and convective zones. Such variation could
indicate the presence of a toroidal magnetic field anchored in this region.

  We found no systematic variation with time of the stratification at the base
of the convection zone. However we can set an upper limit to any fractional
change of the sound speed at the level of $3 \times 10^{-5}$.
\end{abstract}

\section{Introduction}

  Changes in the frequency of the solar $p$-mode oscillations have now been
observed for more than a decade. Such changes affect both the central
frequencies, $\nu_{n \ell}$, and the frequency splittings, $\Delta \nu_{n \ell
m}$ of low degree \citep{bi40}, intermediate degree \citep{bi41,bi43} and very
high degree modes \citep{bi42}. A number of mechanisms have been proposed to
explain these variations on frequency. \citet{bi1} have argued, on the basis
of observations of intermediate degree modes, that the source of the
perturbations must lie near the solar surface. \citet{bi2} and \citet{bi3}
concluded that magnetic fields located near the base of the convection zone,
with strengths significantly lower than $10^6$ G have no observable effect on
$p$-mode frequencies. The stability analysis for magnetic fields by
\citet{bi5} has shown that fields with strengths significantly larger than
$10^5$ G cannot be stored in this region.

The $p$-mode frequency variations track rather well the changes of the
activity strength of the solar cycle with time. It is thus plausible that
these frequency shifts are due to variations of the mean magnetic field near
the photosphere \citep{bi2,bi8}. The influence of thin magnetic fibrils on the
frequency shifts has been investigated by \citet{bi15}. Another possible cause
for these frequency shifts is the presence of sunspots during solar activity
\citep{bi14}. The dominant effect of sunspots on the propagation of acoustic
waves is believed to be the dissipation of the acoustic energy and therefore,
should decrease their amplitudes and lifetimes \citep{bi44,bi42}, while it has
been observed that frequencies increase with increased magnetic activity.

  In the work presented here, we explore a mechanism first suggested by
\citet{bi6}, in which the sound speed perturbation associated to the observed
changes of the photospheric latitudinal temperature distribution might be
responsible for the frequency shifts seen during the solar cycle. Changes in
the temperature distribution are themselves due to the heat transport through
the convection zone induced by the solar dynamo.

  Let us suppose that there is a magnetic field anchored below the base of the
convection zone. To maintain pressure equilibrium, the gas pressure and thus
the density inside the magnetized region must be lower than in its
surroundings. The magnetized fluid will thus experience a larger radiative
heating and therefore the temperature at the top of the magnetized region will
increase. This will also induce a change in the temperature gradient that
could be large enough to make the region above the magnetic field convectively
unstable. In such a scenario, the base of the convection zone would locally
drop, allowing the magnetized fluid to ascend by convective upflows,
transporting excess entropy to the photosphere \citep{kus}.

  Numerical experiments by \citet{kus} have shown that entropy perturbations
in the deep convection zone can produce strongly peaked temperature changes in
regions below $\tau = 1$ that have a substantial acoustic signature (where
$\tau$ is the optical depth). They have also shown that the thermal
perturbations that account for the solar acoustic variability are consistent
with the observed solar irradiance and luminosity changes that occur during
the 11 year solar cycle.

  Luminosity changes, even if no larger than $0.1 \%$, must come from the
release of energy stored somewhere in the solar interior and must be
accompanied by a change in the solar radius. The ratio between relative
luminosity and radius changes, hereafter $W$, can help estimate the location
of the region where this energy is stored \citep[and references
therein]{go2000}. Theoretical calculations indicate that $W$ increases when
increasing the depth of the source of the variations in luminosity. For
instance $W \approx 2 \cdot 10^{-4}$ if the source is located in the outer
layers of the convection zone, while $W \approx 0.5$ if the source is located
in the solar core. Unfortunately there is a large scatter in the observed
values of $W$. Indeed, recent measurements of $W$ range from 0.021 as
estimated by \citet{bi45} and \citet{bi46} to an upper limit of 0.08 derived
by \citet{emi}.

   \citet{ku} estimated that a $0.1 \%$ luminosity perturbation integrated
over a solar cycle corresponds to about $10^{39}$ erg. If this energy
originates in the tachocline and if the tachocline thickness is $0.05\,\Rsun$,
the associated relative variation in sound speed at that depth would be on the
order of $\delta c / c \approx 10^{-5}$ or $10^{-6}$. Fractional changes in
sound speed as small as $10^{-4}$ are easily accessible by helioseismic
inversion techniques. Some attempts to find solar-cycle variations of the
sound speed asphericity and the latitude-averaged sound speed have been
carried out using MDI and/or GONG data
\citep{bi45,bi41,bi47,bi48,bi49}. However, none of them found any systematic
variation of the solar structure at the base of the convection zone that could
be associated with the presence of a local toroidal magnetic field.

We have extended the previous analysis to the latest data available, including
LOWL data. Only common modes to all data sets were used, in an attempt to
obtain comparable and significant results to all instruments. This way, we can
give a robust upper limit on the temporal variations of the solar internal
stratification during the period 1994-2001.
 
\section{Inversion Technique}

  The inversion for solar structure, in particular sound speed $c$ and density
$\rho$, are commonly based on the linearization of the equations of stellar
oscillations around a reference model \citep{goug5,dzi2,goug6}. The
differences of the structural profile between the actual sun and a model are
linearly related to differences between the observed frequencies and those
calculated using that model. This relation is obtained using a variational
formulation for the frequencies of adiabatic oscillations. A general relation
for frequency differences is given by

\begin{eqnarray}
\frac { \delta \nu_{nl}}{ \nu_{nl}} & = & 
  \int_0^{\Rsun} { 
   \left[ K^{nl}_{c,\rho}(r) \frac{\delta c}{c}(r)
        + K^{nl}_{\rho,c}(r) \frac{\delta \rho}{\rho}(r)
  \right] dr}  \label{test2} \\  
   ~&~& + \ {\cal E}^{-1}_{nl}F(\nu) + \epsilon_{nl} \nonumber
\label{test3}
\end{eqnarray}
where $\delta \nu_{nl}$ are the frequency differences between the actual sun
and the model for the mode with radial order $n$ and degree $l$, and
$\epsilon_{nl}$ the corresponding relative error.  The sensitivity functions,
or kernels, $K^{nl}_{c,\rho}(r)$ and $K^{nl}_{\rho,c}(r)$ are known functions,
that relate the changes in frequency to the changes in the model.  The
functions $\delta c /c$ and $\delta \rho / \rho$ are the unknown parameters to
be inverted, \ie: the relative difference in the sound speed and the density
respectively, and $\Rsun$ is the solar radius.  The term ${\cal
E}^{-1}_{nl}F(\nu_{nl})$ in Eq.~\ref{test2} is introduced to take into
account the so-called surface uncertainties; these include the dynamical
effects of convection on the oscillation equations, as well as non-adiabatic
processes in the near-surface layers \citep[see][and references therein]{dzi2}.

  Following standard procedures, we represent $F(\nu_{nl})$ as a
Legendre polynomial expansion. ${\cal E}_{nl}$ is the inertia of the mode,
normalized by the inertia that a radial mode of the same frequency would have
\citep[for more details, see][]{goug}.

  If we take $\delta \nu_{nl}$ as the differences in the observed
frequencies at two different epochs, rather than the differences in frequency
between the actual sun and the model, $\delta c /c$ and $\delta \rho / \rho$
represent the variation with time of the sun's internal structure, as long as
our underlying theoretical model is very close to the actual sun.

  The inverse problem defined by Eq.~\ref{test2} is well known to be an
ill-posed problem \citep{Thompson:1995}, whose solution is not unique. It can
be solved using inversion methodologies that can be classified in two
different techniques: the regularized least-squares methods \citep[RLS,
see][]{bi23} and the optimal localized average methods
\citep[OLA,][]{Back}. Both methods compute an estimate of the solution at a
target location from a linear combinations of the observables, given a mesh of
target locations.
  We have developed a variant of the RLS technique, that we call the optimal
mesh distribution (OMD), that optimizes the mesh of target locations to avoid
undesired high-frequency oscillations of the solution. This optimization is
achieved by computing {\em a priori} the spatial resolution of the solution
from the set of available observables and their uncertainties
\citep{bi12}. The smoothing function is itself defined also from the spatial
resolution analysis and it is weighted differently for each radial point. This
method ensures that the smoothing constraint is properly applied over the
optimal mesh.

\section{Observational Data}

  The observational data consist of mode frequencies computed from time series
spanning different epochs and observed with different instruments. Namely, 57
sets based on 108-day-long time series derived from the GONG instruments
\citep{bi9} and spanning May 1995 to February 2001; 27 sets based on
72-day-long time series derived from the MDI instrument \citep{bi11} and
spanning May 1996 to November 2001; and 6 sets based on 1-year-long time
series derived from the LOWL instrument \citep{jim,tom} and spanning 1994 to
1999.

  In order to use consistent data sets, only the modes common to all the sets
for a given instrument were taken into account. As a consequence, the low
degree modes $(l<13)$ present in some GONG data sets had to be rejected. Also
this selection reduces the number of MDI and LOWL modes by $30 \%$ and $4 \%$
respectively.

 The MDI and LOWL sets were further reduced to only include the modes common
to both instruments. This was not done with the GONG data set due to the small
amount of common modes present. Finally, and again for consistency, we
deliberately restricted range of degrees we included to correspond to the
highest degree available in the LOWL data set (\ie, $l \le 100$).

  For each instrument and for each mode we computed the temporal frequency
average. We subsequently subtracted the respective averaged frequencies from
each set, leaving us with frequency changes with respect to this temporal
average as a function of epoch.  For the GONG and MDI sets, we also computed
averages corresponding to 1-year-long epochs. Such averaging reduced the
scatter of the data while producing data sets comparable to the LOWL sets.

\section{Results}

\begin{figure}[!ht]
\vspace*{427.50pt} 
\includegraphics{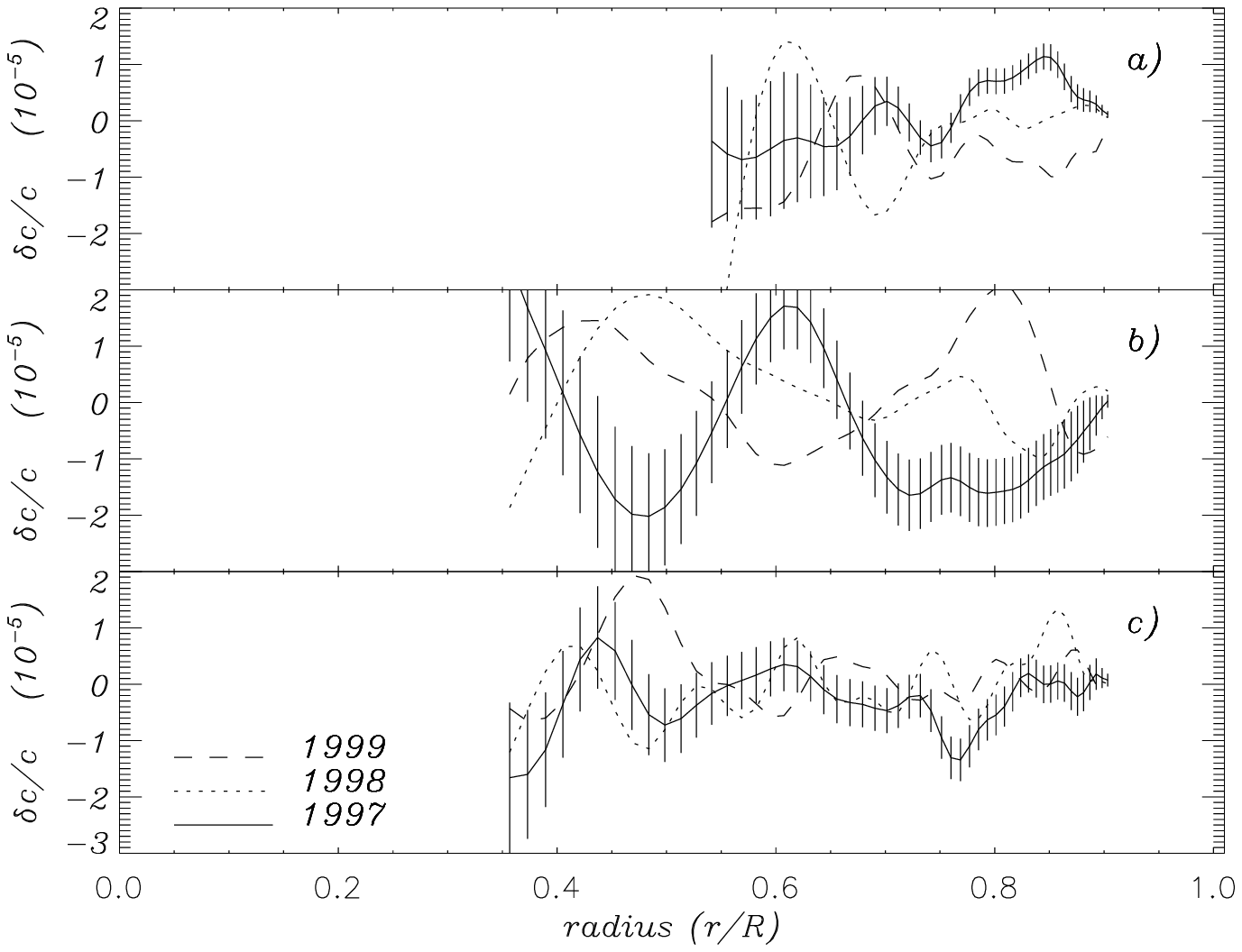} 
\caption{Relative sound speed differences inferred using 1-year-long averaged
GONG (panel a), LOWL (panel b) and MDI (panel c) data, shown for clarity only for the years 1997, 1998 and 1999. The
solutions are shown only where the inversions are reliable. For clarity, the
error bars are only shown for the inversion of the 1997 data; the error bars
of the other results are almost identical.}
\label{fig1}
\end{figure}

\begin{figure}[!ht]
\vspace*{427.50pt} 
\includegraphics{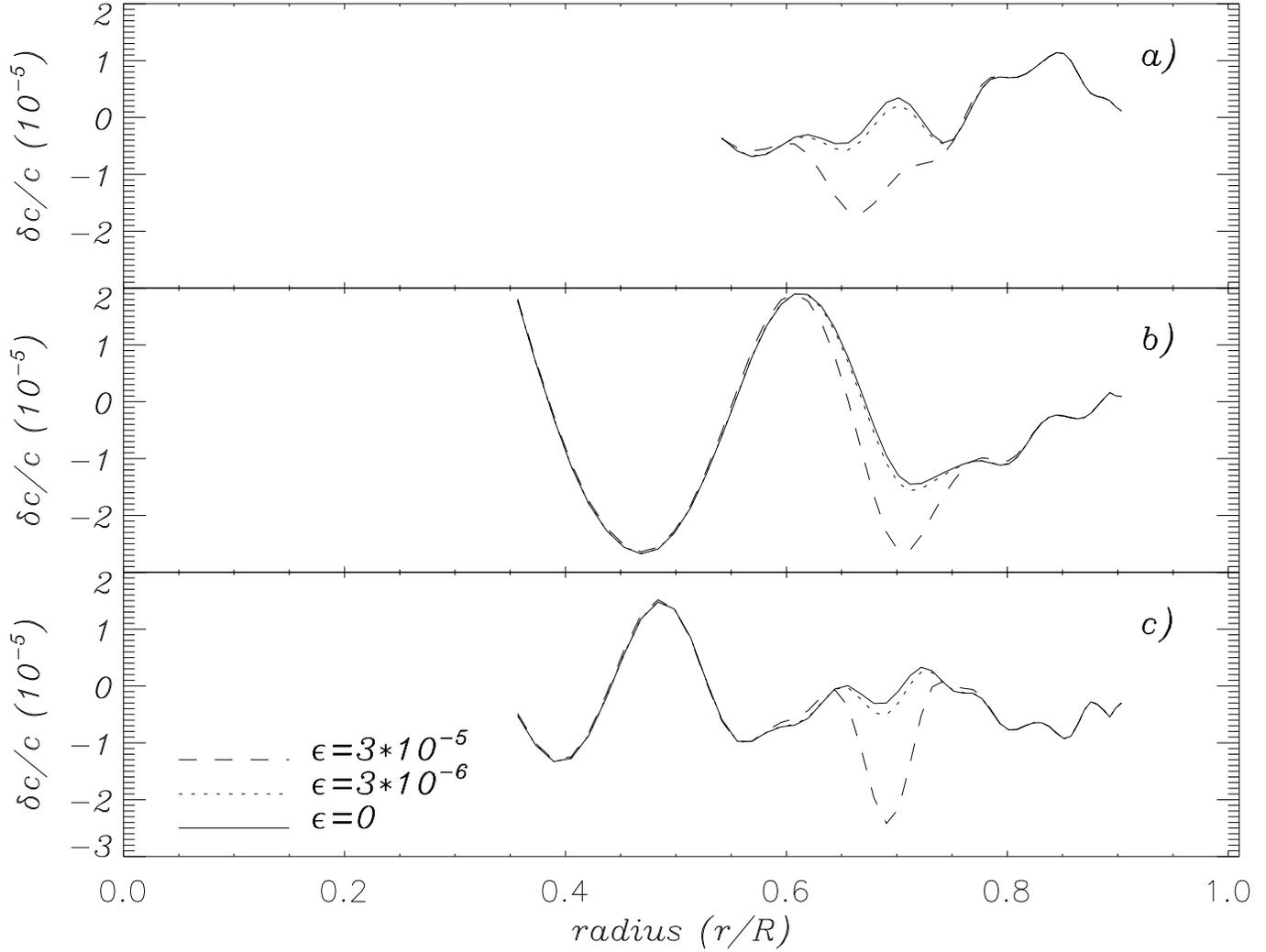}
\caption{Relative sound speed differences inferred from the 1996 averaged GONG (panel a), LOWL (panel b) and MDI (panel c) data sets. For each panel, the solid line shows the inversion similar to those presented in
Fig.\ 1, whereas the other two profiles correspond to inverted profiles from the same
mode set but after the injection of a frequency change corresponding to a
small perturbation in the sound speed between $0.68$ and $0.70\,\Rsun$, of
amplitude $\varepsilon = 3 \times 10^{-6}$ and $ 3 \times 10^{-5}$
respectively. For clarity, error bars have not been included, although they
are very similar to those shown in Figure~\ref{fig1}.}
%
%
\label{fig2}
\end{figure}

  Figure~\ref{fig1} shows the relative change of the sound speed as a function
of radius inferred from 1-year-long MDI, GONG and LOWL sets. These profiles show no
significant changes at the level of a few times $10^{-5}$. The precision
and resolution of the inversion is good enough to detect small variations of
the stratification at the base of the convection zone. This is demonstrated in
Fig.~\ref{fig2}, where we show the sound speed profiles inferred from the 1996
averaged MDI, GONG and LOWL data sets as well as sound speed profiles obtained by inverting
the same mode sets, but after injecting frequency changes that result from a
perturbation in the sound speed (as small as $3 \times 10^{-6}$ and $3 \times
10^{-5}$) between $0.68$ and $0.70\,\Rsun$. This figure indicates that
a perturbation of the sound speed at the base of the convection zone on the
order of, or slightly smaller than $5 \times 10^{-5}$, can be detected
with the current precision resulting from 1-year-long
time-series. Perturbations on the order of $10^{-6}$ fall in the noise level
of our inversions.

\begin{figure}[!ht]
\vspace*{427.50pt} 
\includegraphics{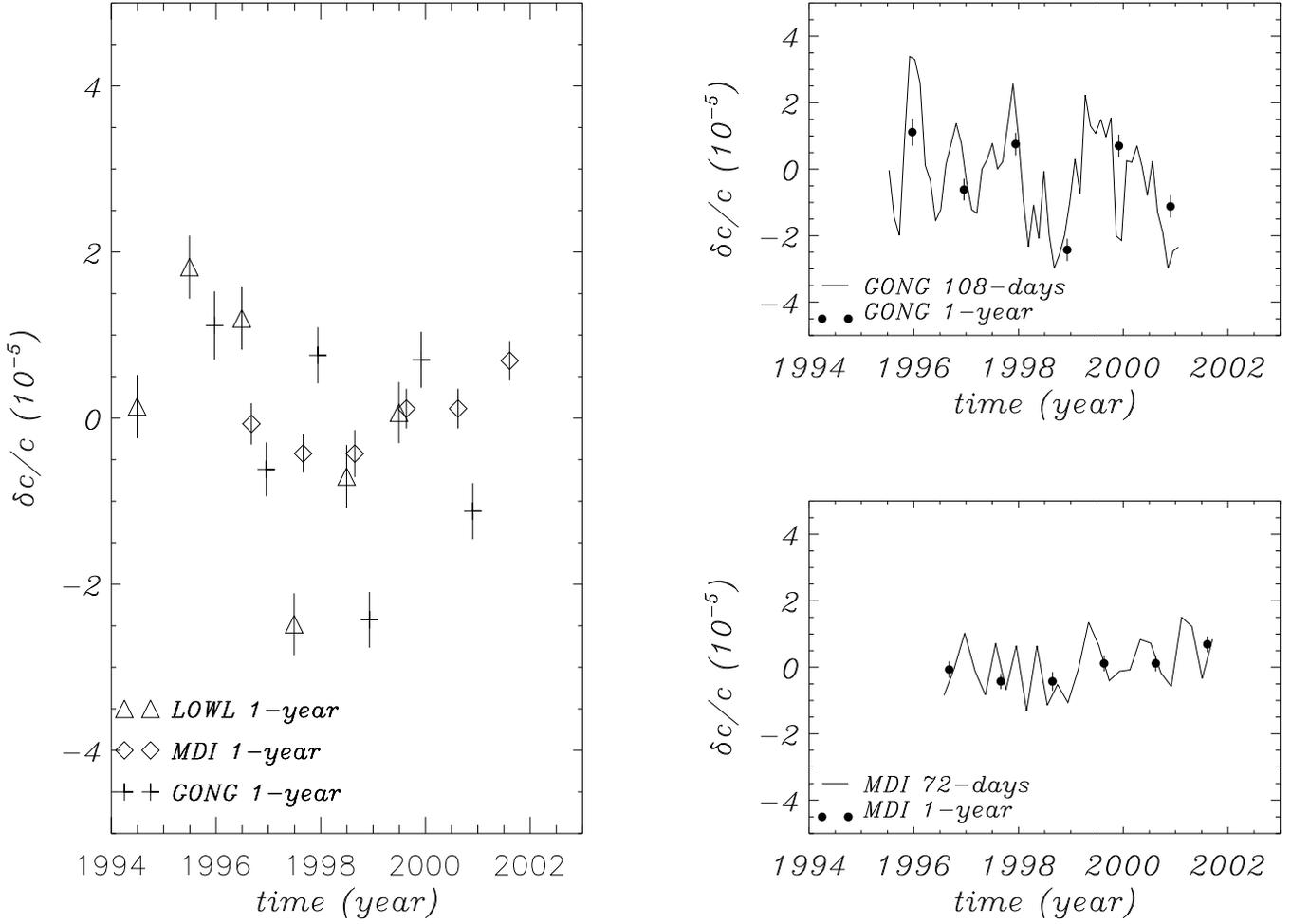}
\caption{Temporal variations of the average of $\delta c / c$ at
$r_0=0.69\,\Rsun$ computed over the radial points that lie within the interval
$0.69 \le r/\Rsun \le 0.72$. Left panel shows the results for LOWL, MDI and
GONG 1-year-long data sets. Right panels compare results for the 108-day-long
and 1-year-long GONG sets and the 72-day-long and 1-year-long MDI sets.}
\label{fig3}
\end{figure}

  In an attempt to find temporal variations of the solar stratification at the
base of the convection zone, we computed the mean value of ${\delta c}/{c}$ in
the radial interval $0.69 \le r/\Rsun \le 0.72$. This interval contains not
only the base of the convection zone, $r \approx 0.7133\,\Rsun$ \citep{bi49},
but also the tachocline, $r \approx 0.691\,\Rsun$ \citep{bi50}, both closely
related to the toroidal magnetic field responsible of the solar cycle. The
resulting values are shown as a function of time in Fig.~\ref{fig3}, for the
inversions based on 1-year-long sets for all three instruments (GONG, MDI and
LOWL), as well as on the GONG 108-day-long and MDI 72-day-long sets.

  Inversion profiles inferred from any linear inversion technique always
correspond to the convolution of the underlying solution by the resolution
kernel \citep{Thompson:1995}. Therefore, even if the mode set used in a
sequence of inversions remains identical, the resolution kernels will, at some
level, change with time since the uncertainties change with time. Such
variation could produce an {\em apparent} temporal behavior of the inferred
profiles that does not correspond to a {\em real} variation of the underlying
{\em true} solution. To quantify this effect, we computed the averaging
kernels at $r=0.69\,\Rsun$ for all the inversions based on 1-year-long data
sets. These were convolved with an artificial sound speed perturbation of the
form:
\begin{equation}
 \frac{\delta c}{c} = \left\{  
 \begin{array}{ll}    
    3 \times 10^{-5} &  \mbox{for $0.67 \le r/\Rsun \le 0.71 $} \\
    5 \times 10^{-5} &  \mbox{for $0.91 \le r/\Rsun \le 0.93 $} \\
    0                &  \mbox{otherwise}
\end{array} \right.  
\end{equation}
where the sound speed perturbation centered at $0.92\,\Rsun$ attempts to
reproduce the results found by \citet{bi43,bi48}, while a second perturbation
at the base of the convection zone is also introduced. The convolutions
$q_{r_o}(t)$ are then calculated in the following way:

\begin{equation}
  q_{r_o}(t) = \int K(r,r_o,t)\, \frac{\delta c(r)}{c(r)}\ dr
\end{equation}

\begin{figure}[!ht]
\vspace*{427.50pt} 
\includegraphics{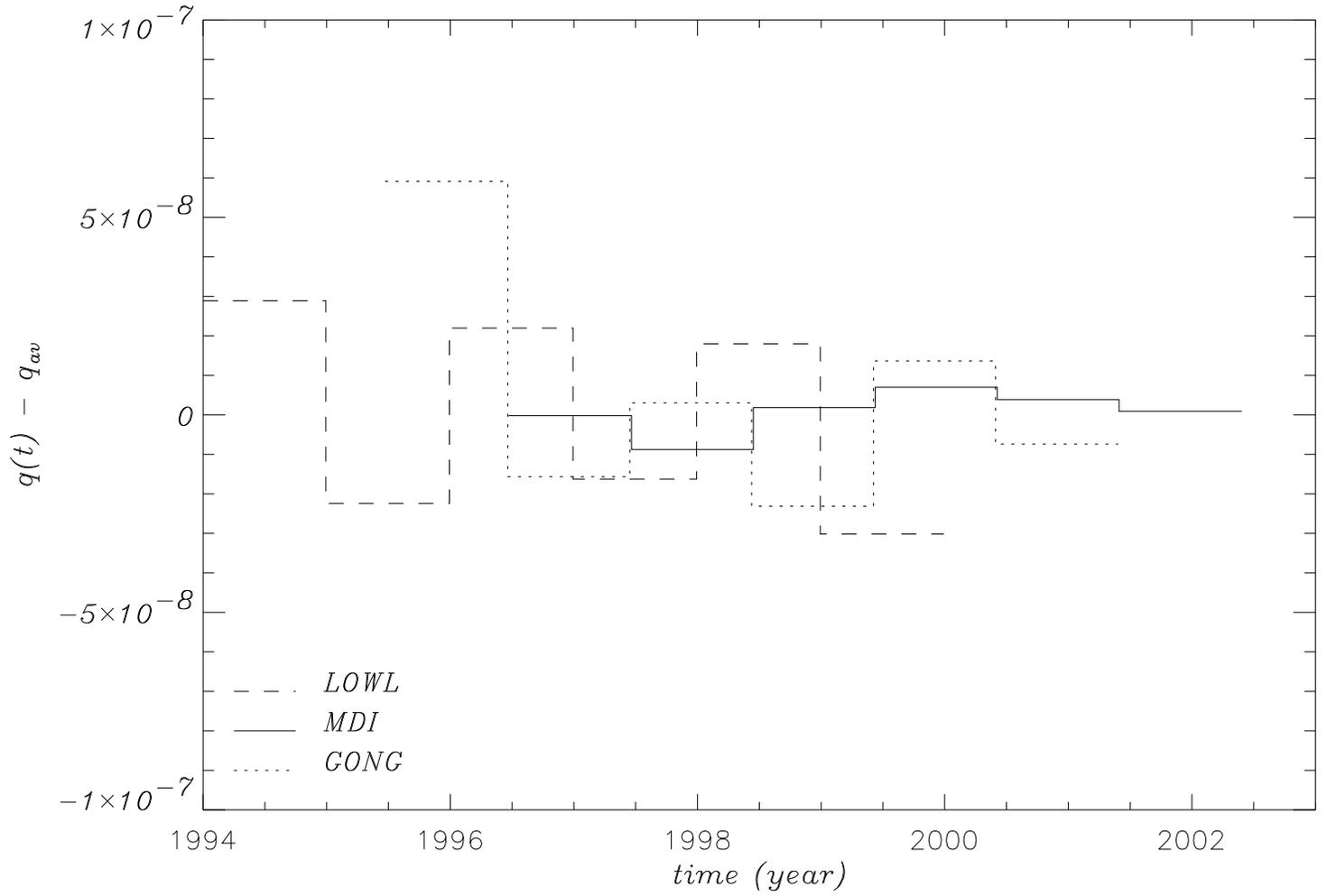} 
\caption{Contribution of the time-varying averaging kernel to spurious sound
speed profile changes with time; see text for details.}
\label{fig4}
\end{figure}

The resulting values of $q_{r_o}(t)$, at $r_o=0.69\,\Rsun$, relative to the
average $q_{av}$ of all the convolutions calculated for every instrument, are
shown in Fig.~\ref{fig4}. This figure demonstrates that for the data from all
three instruments the effect of the changes in the observed uncertainties on
the inverted profiles is negligible, corresponding to levels well below
$10^{-7}$.

\begin{figure}[!ht]
\vspace*{427.50pt} 
\includegraphics{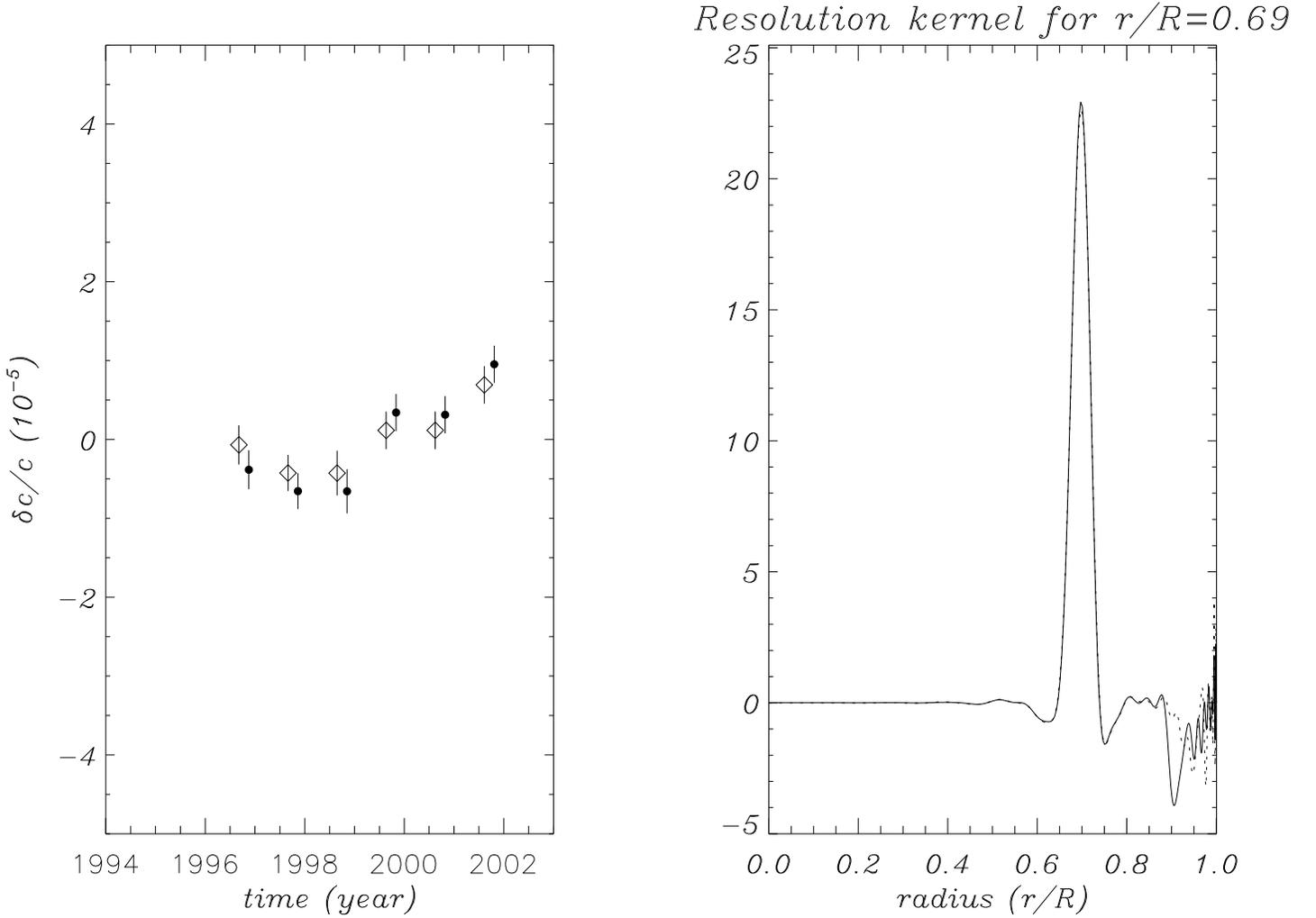} 
\caption{Left panel: temporal variations of the average of $\delta c / c$ over
the radial points that lie within the interval $0.69 \le r/\Rsun \le
0.72$. Results correspond to 1-year-long MDI data with degrees expanding up to
$l=100$ (shown as diamonds) and $l=150$ (shown as filled circles). Right
panel: averaging kernels corresponding to the inversion of the 1996 data at
$r=0.69\,\Rsun$. The solid line corresponds to the averaging kernel obtained
from the inversion of data with degrees up to $l=100$, while the dotted lines
represents the averaging kernel obtained from the inversion of data with
degrees up to $l=150$.}
\label{fig5}
\end{figure}

The averaging kernel corresponding to the solution at $r=0.69\,\Rsun$ obtained
from the 1997 MDI data set is shown in the right panel of
Fig.~\ref{fig5}. This is well located and indicates that our radial resolution
corresponds to $0.04\,\Rsun$. We should also point out that the averaging
kernels have an important negative non-local contribution near the surface, a
feature that affect at some level the solution at the base of the convection
zone. The effect of this non-local component of the averaging kernels was
quantified, in the case of MDI and GONG data, by performing inversions with
data sets expanded to higher degrees, \ie, up to $l=150$. The averaging kernel
at $r=0.69\,\Rsun$ obtained from the MDI 1997 data shows a substantial
reduction of the negative component located at $ r \approx 0.90\,\Rsun$ when
including higher degree modes. But the temporal behavior of the
stratification at the base of the convection zone do not significantly differ
when the MDI data sets are expanded from $l \le 100$ to $l \le 150$, as
illustrated in the left panel of Fig.~\ref{fig5}.

Therefore, by including the data from all three instruments, and after
assessing the effects of temporal changes of the resolution of the solutions,
we can safely conclude that there is not significant systematic variations of
the stratification at the base of the convection zone at the level of $3
\times 10^{-5}$, and that this upper limit is constrained by the scatter
present in the data.

\citet{vor} analyzed MDI data spanning from 1996 to 2000 and found systematic
variations of the radial solar stratification with time, expressed as relative
changes of radius of $2 \times 10^{-5}$. Our results obtained with MDI data
are in good agreement with those found by \citet{vor}, in the sense that there
is a systematic variation in the relative sound speed difference that is well
correlated to the magnetic activity in the Sun. The maximum in solar activity corresponds to the maximum variation in sound speed. Since these results are
not seen when analyzing LOWL and GONG data they should still be taken with
some degree of scepticism.

\section{Acknowledgments}

  The Solar Oscillations Investigation - Michelson Doppler Imager project on
SOHO is supported by NASA grant NAS5--3077 at Stanford University. SOHO is a
project of international cooperation between ESA and NASA.

  The GONG project is funded by the National Science Foundation through the
National Solar Observatory, a division of the National Optical Astronomy
Observatories, which is operated under a cooperative agreement between the
Association of Universities for Research in Astronomy and the NSF.

The LOWL instrument has been operated by the High 
Altitude Observatory of the National Center for Atmospheric Research which is 
supported by the National Science Foundation.

  This work was partially supported by NASA -- Stanford contract PR--6333
and by NSF grant AST--95--2177.

\newpage
\onecolumn

\newpage

\newpage

\newpage

\newpage

\end{document}